\documentclass[11pt,a4paper]{article}
\usepackage[hyperref]{emnlp-ijcnlp-2019}
\usepackage{times}
\usepackage{latexsym}

\usepackage{url}
\usepackage{graphicx}
\usepackage{booktabs}
\usepackage{amsmath}
\usepackage{multirow}
\usepackage{amsfonts}

\aclfinalcopy % Uncomment this line for the final submission

\newcommand{\start}[1]{\vspace{1.8mm}\noindent{{\bf #1}}}
\newcommand{\ie}{\textit{i.e.}} % that is
\newcommand{\eg}{\textit{e.g.}} % for example

\title{Hierarchical Text Classification with Reinforced Label Assignment}

\author{Yuning Mao$^{1}$, Jingjing Tian$^{2}$, Jiawei Han$^1$, Xiang Ren$^3$ \\
$^1$Department of Computer Science, University of Illinois at Urbana-Champaign, IL, USA \\
$^3$Department of Computer Science, Peking University, Beijing, China\\
$^3$Department of Computer Science, University of Southern California, CA, USA\\
$^1$\{yuningm2, hanj\}@illinois.edu $\quad$
  $^2$tianjj97@pku.edu.cn $\quad$ $^3$xiangren@usc.edu
}

\date{}

\begin{document}
\maketitle
\begin{abstract}
While existing hierarchical text classification (HTC) methods attempt to capture label hierarchies for model training, they either make local decisions regarding each label or completely ignore the hierarchy information during inference. 
To solve the mismatch between training and inference as well as modeling label dependencies in a more principled way, 
we formulate HTC as a Markov decision process and propose to learn a \textbf{L}abel \textbf{A}ssignment \textbf{P}olicy via deep reinforcement learning to determine \textit{where to place} an object and \textit{when to stop} the assignment process. 
The proposed method, \textbf{HiLAP}, explores the hierarchy during both training and inference time in a \textit{consistent} manner and makes \textit{inter-dependent} decisions.
As a general framework, HiLAP can incorporate different neural encoders as \textit{base models} for end-to-end training.
Experiments on five public datasets and four base models show that HiLAP yields an average improvement of 33.4\% in Macro-F1 over flat classifiers and outperforms state-of-the-art HTC methods by a large margin.\footnote{\scriptsize Data and code can be found at \url{https://github.com/morningmoni/HiLAP}.}
\end{abstract}

\section{Introduction}

In recent years there has been a surge of interest in leveraging hierarchies (taxonomies) to organize objects (\eg, documents), leading to the development of hierarchical text classification (HTC)---a task that aims to predict for an object multiple appropriate labels in a given label hierarchy, which together constitute a sub-tree. HTC methods have found a wide range of applications such as question answering~\citep{Qu2012AnEO}, online advertising~\citep{agrawal2013multi}, and scientific literature organization~\citep{peng2016deepmesh}.
In contrast to ``flat'' classification, the key challenges of HTC lie in modeling the large-scale, imbalanced, and in particular, structured label space.

\begin{figure}[t]
    \centering
    \includegraphics[width=1.0\linewidth]{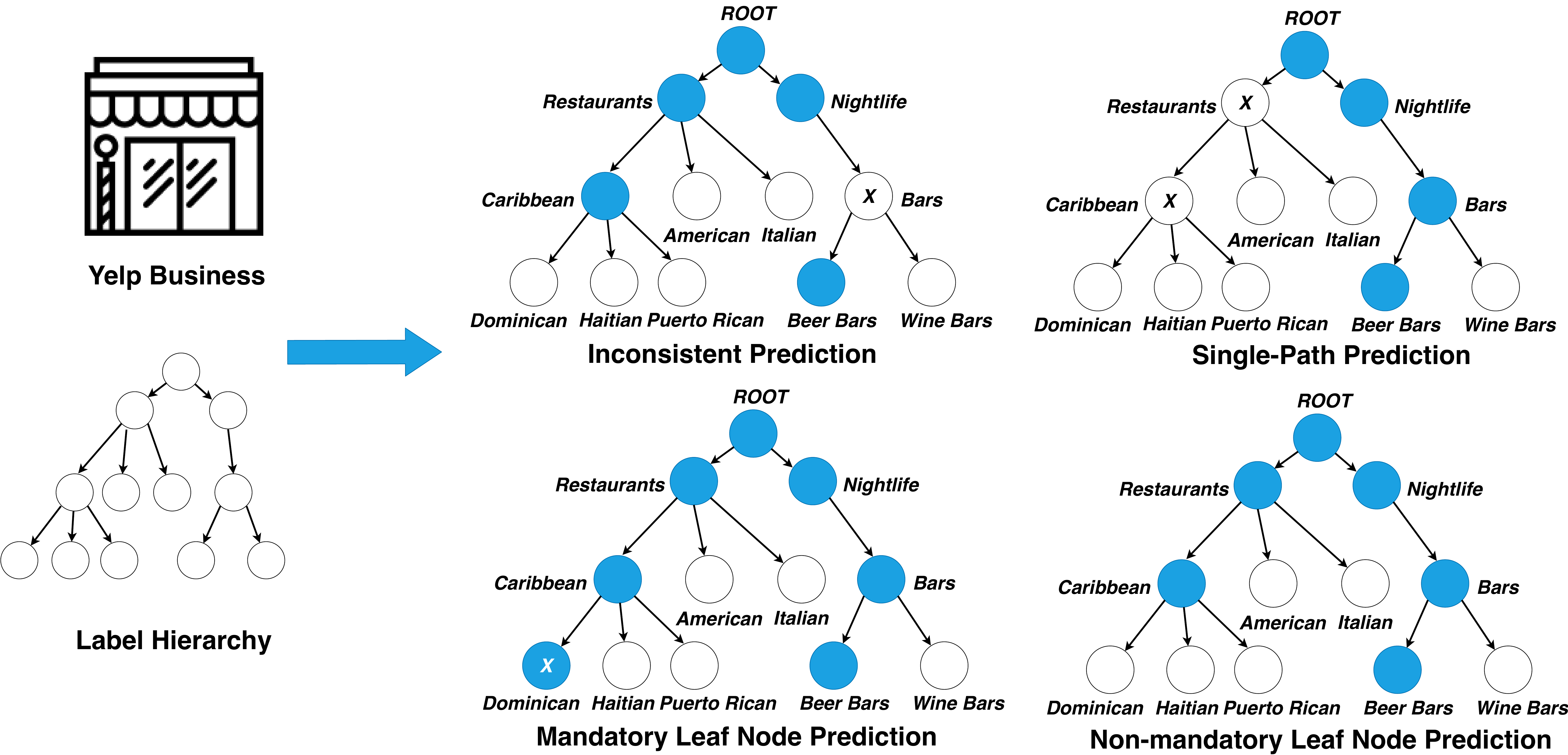}
     \caption{We aim at \textbf{consistent, multi-path, and non-mandatory leaf node prediction}. For a Caribbean restaurant with a beer bar, inconsistent prediction may place it to node ``Beer Bars'' but not ``Bars'', which contradicts with each other; Single-path prediction may only recognize that it is a beer bar; Mandatory leaf node prediction would have to assign a leaf node ``Dominican'' even if the nation of the cuisine is uncertain.}
    \label{fig:examples}
    \vspace*{-.1cm}
\end{figure}

\begin{figure*}[ht]
    \centering
    \includegraphics[width=0.95\linewidth]{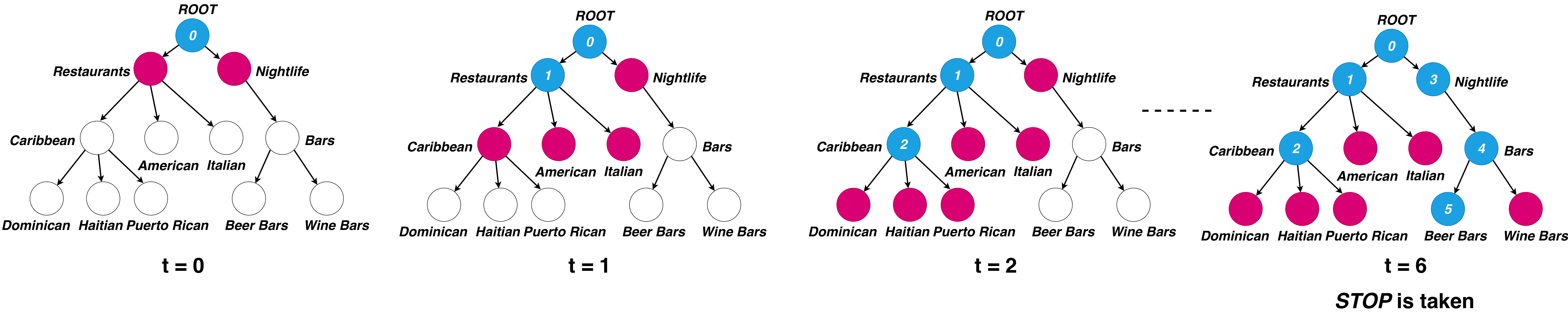}
    \vspace*{-.15cm}
     \caption{\textbf{An illustrative example of the label assignment policy.} At $t=0$, $x_i$ is placed at the root label and the policy would decide if $x_i$ should be placed to its two children (\textit{red}). At $t=1$, $x_i$ is placed at label ``Restaurants'', which adds its three children as the candidates. At $t=6$, the \textit{stop} action is taken and the label assignment is thus terminated. We then take all the labels where $x_i$ has been placed (\textit{blue}) as $x_i$'s labels.}
    \label{fig:example}
    \vspace*{-.2cm}
\end{figure*}

Based on how the hierarchy is explored, HTC methods can be summarized into \textit{flat}, \textit{local}, and \textit{global} approaches~\citep{silla2011survey}.
Flat approaches~\cite{hayete2005gotrees,johnson2014effective} assume all the labels in the given hierarchy are independent. 
Some predict labels at the leaf nodes and heuristically add their ancestor labels, which is problematic as the labels of some objects may not be at the leaf nodes (\textit{non-mandatory leaf node prediction}, see Fig.~\ref{fig:examples}) and all the non-leaf nodes are completely neglected.
Some simply ignore the hierarchy and perform standard multi-label classification, in which \textit{label inconsistencies} (one label is predicted positive but its ancestors are not) may occur and post-processing is needed to correct such contradictions. 
Local approaches~\cite{koller1997hierarchically,cesa2006hierarchical} train a set of local classifiers that function \textit{independently} and predictions are usually made in a top-down order: one node is visited if and only if its ancestors have been predicted positive.
One critical issue is that the number of local classifiers depends on the size of the label hierarchy, making local approaches infeasible to scale.

Global approaches use one single classifier and model the label hierarchy more explicitly.
Traditional global approaches~\citep{wang2001hierarchical,silla2009global} are largely based on specific flat models and often make unrealistic assumptions~\cite{cai2004hierarchical} as in flat approaches.
Recent neural approaches~\citep{kim2014convolutional,yang2016hierarchical} mainly focus on flat classification while their performance in HTC is relatively less studied.
Even if the classification is supposed to be hierarchical, prior work~\citep{gopal2013recursive,johnson2014effective,peng2018large} still make \textit{flat} and \textit{independent} predictions or utilize simple constraints without considering the holistic quality of label assignment.
One recent framework~\citep{wehrmann2018hierarchical} attempts to leverage both local and global information but it uses static features as input and its inference process is still flat.

\begin{figure*}[ht]
    \centering
    \includegraphics[width=0.92\linewidth]{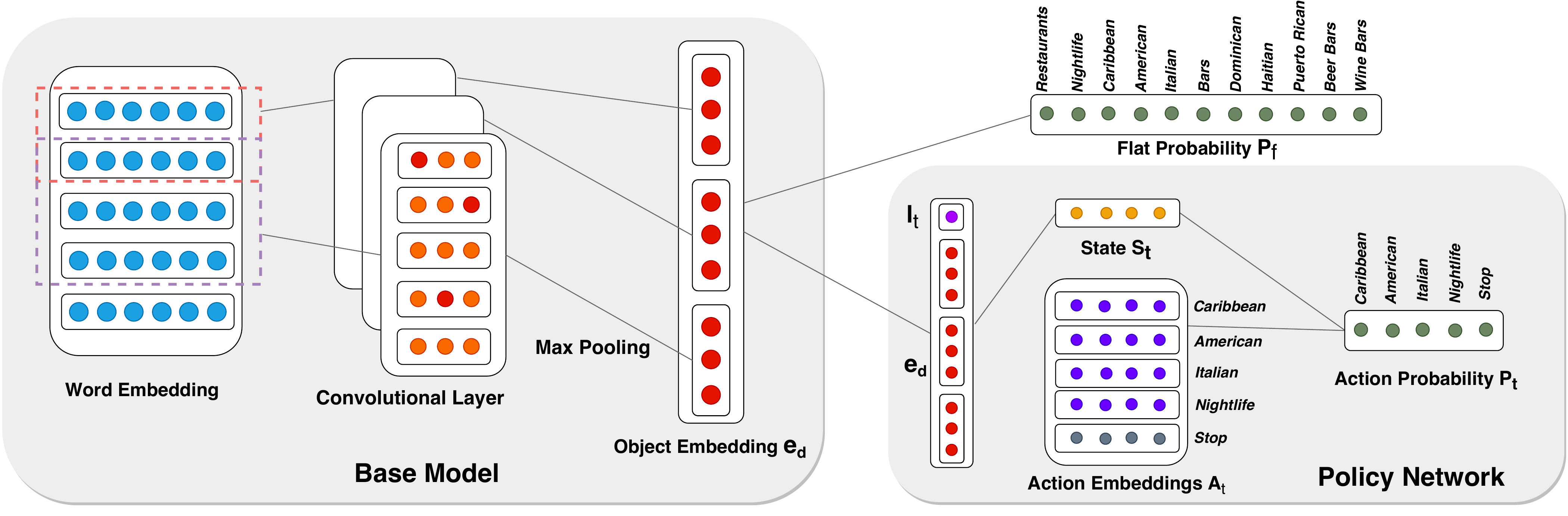}
     \caption{\textbf{The architecture of the proposed framework HiLAP}. One CNN model~\citep{kim2014convolutional} is used as the base model for illustration. The object embedding $\textbf{e}_d$ generated by the base model is combined with the embedding of currently assigned label $\textbf{l}_t$ and used as the state representation $\textbf{s}_t$, based on which actions are taken by the policy network. The time corresponds to $t=1$ in Fig.~\ref{fig:example}.}
    \label{fig:net_arch}
    \vspace*{-.1cm}
\end{figure*}

In this paper, we formulate HTC as a Markov decision process to better capture \textit{label dependencies} and measure the \textit{holistic quality} of label assignment.
We present HiLAP, a global framework that learns a label assignment policy to determine \textit{where to place} the objects and \textit{when to stop} the assignment process.
HiLAP explores the label hierarchy during both training and inference in a consistent manner, which alleviates the exposure bias often found in prior local and global approaches.
By learning when to stop, HiLAP is more flexible than approaches that only support mandatory leaf node prediction or require thresholding.
In addition, HiLAP supports multi-path prediction and its predictions of one object on different paths are \textit{inter-dependent}, which not only guarantees \textit{label consistency} but matches the nature of HTC.
Furthermore, HiLAP estimates the holistic quality of all the labels assigned to one object via reinforcement learning instead of evaluating each label independently via maximum likelihood as in prior studies.
To summarize, HiLAP achieves better effectiveness compared to flat and local approaches as it examines the label hierarchy during both training and inference.
HiLAP has more flexibility and generalization capacity than previous global approaches in that it has no constraints on the structure of the hierarchy or the labels of the objects~\citep{cai2004hierarchical}, generalizes to neural representation learning models~\citep{gopal2013recursive}, and makes inter-dependent predictions while ensuring label consistency~\citep{wehrmann2018hierarchical,peng2018large}.

HiLAP can be combined with various neural encoding models and trained in an end-to-end fashion.
In our experiments, we select four representative encoding models as the \textit{base models} to evaluate the effectiveness of HiLAP.
Experimental results on five public datasets from different domains show that combining the base models with HiLAP yields an average performance improvement of 33.4\% in Macro-F1 over corresponding flat classifiers and outperforms state-of-the-art HTC methods by a large margin.
In particular, ablation study shows that HiLAP is especially beneficial to those unpopular labels at the bottom levels.

\section{Hierarchical Label Assignment}
\label{subsec_hla}

\subsection{Overview}

\noindent
\textbf{Problem Formulation.}
We define a label hierarchy $H = (L, E)$ as a tree or DAG (directed acyclic graph)-structured hierarchy with a set of nodes (labels) $L$ and a set of edges $E$ indicating the parent-child relation between the labels.
Taking a set of objects $\mathcal{X} = \{x_1, x_2, ..., x_N\}$ and their labels $\mathcal{L} = \{L_1, L_2, ..., L_N\}$ as input, we aim to learn a label assignment policy $\mathcal{P}$ to place each object $x_i$ to its labels $L_i$ on the label hierarchy $H$.
The label assignment is supposed to be \textit{consistent, multi-path, and non-mandatory leaf node prediction} (refer to Figs.~\ref{fig:examples} and~\ref{fig:example}).
We define one \textit{base model} $\mathcal{B}$ as a mapping $f$ that converts raw object $x_i$ to a finite dimensional vector, \ie, the object embedding $\textbf{e}_d \in \mathbb{R}^D$.
$\mathcal{B}$ can be any neural representation learning model and its output $\textbf{e}_d$ is used as the input of $\mathcal{P}$ for policy learning.
The major challenge, compared to standard classification setup, is that we need to model $E$, \ie, the relation between labels.

\smallskip
\noindent
\textbf{Our Framework.}
Prior studies either have a mismatch between training and inference as different routines are followed in the two phases, or compute losses with respect to each individual label and make flat predictions during inference time.
In contrast, we learn a policy that (1) makes consistent, inter-dependent predictions by traversing the label hierarchy and maintaining state representation; (2) measures the holistic quality of label assignment via reinforcement learning.
Specifically, the policy $\mathcal{P}$ puts $x_i$ at the root label in the beginning. At each time step, $\mathcal{P}$ decides which label $x_i$ should be further placed to, among all the \textit{children} labels of where $x_i$ has been placed, until a special \textit{stop} action is taken.
An illustration of how HiLAP labels one object is shown in Fig.~\ref{fig:example} and the overall architecture of HiLAP is shown in Fig.~\ref{fig:net_arch}.

\subsection{Reinforcement Learning for Hierarchical Label Assignment}
\label{sec_rl4taxo}
We describe the details of policy learning including its actions, rewards, states, and the policy network in this section.
We formulate HTC as a Markov decision process (MDP): at each time step, the agent observes current state, takes an action, and receives a reward.
The end goal is to train a policy network to determine where to place the objects and when to stop.

\start{Actions.}
Specifically, we regard the process of placing an object $x_i$ to the right positions on the label hierarchy as making a sequence of actions, where an action $a_t$ at time step $t$ is to select one label $l_{t+1}$ from the action space $\mathcal{A}_t$ and place $x_i$ to that label $l_{t+1}$.
We denote the children of label $l_t$ as $\mathcal{C}(l_t)$.
At the beginning of each episode, $x_i$ is placed at the root label $l_0$ and the action space $\mathcal{A}_0 = \mathcal{C}(l_0)$, \ie, all the labels at level 1.
When $x_i$ is placed at another label $l_1$, its children $\mathcal{C}(l_1)$ are then added to the action space $\mathcal{A}_1$ while $l_1$ itself is removed.
In addition, one \textit{stop} action with embedding $\textbf{e}_{\text{stop}} \in \mathbb{R}^C$ is included in the action space so that the model can automatically learn when to stop placing object $x_i$ to new labels.
Intuitively, when the confidence of placing $x_i$ to another label is lower than the \textit{stop} action, the label assignment process would be terminated.

In short, the action space $\mathcal{A}_t$ consists of all the \textit{unvisited} children labels of where the object $x_i$ \textit{has been placed} and the \textit{stop} action.
One distinction of HiLAP is that it takes the inter-dependencies of labels across different paths and levels into consideration while previous approaches make independent predictions on different paths.
For example, HiLAP can first place $x_i$ to a label at level 3 if the probability of that label is high and then place it to another label at level 1 on another path.

\start{Rewards.}
The agent receives scalar rewards as feedback for its actions.
Different from existing work where each label of one example\footnote{We use ``example'' and ``object'' interchangeably.} is treated independently, HiLAP measures the quality of all the labels assigned to each example $x_i$ by rewarding the agent with the \textit{Example-based F1} (see Sec.~\ref{subsec:metric} for details of this metric).
Intuitively, the agent would realize how similar the assigned and the ground-truth labels of \textit{one example} are.
Instead of waiting until the end of the label assignment process and comparing the predicted labels with the gold labels, we use reward shaping~\citep{P18-1229}, \ie, giving intermediate rewards at each time step, to accelerate the learning process.
Specifically, we set the reward $r$ of $x_i$ at time step $t$ to be the difference of Example-based F1 scores between current and the last time step: $r_t^{x_i} = \text{F1}_{t}^{x_i} - \text{F1}_{t-1}^{x_i}$.

If current F1 is better than that at the last time step, the reward would be positive, and vice versa.
The cumulative reward from current time step to the end of an episode would cancel the intermediate rewards and thus reflect whether the current action improves the holistic label assignment or not.
As a result, the learned policy would not focus on the current placement but have a long-term view that takes following actions into account.

\start{States and Policy Network.}
We parameterize action $a_t$ by a policy network $\pi(\textbf{a}\ |\ \textbf{s}; \textbf{W})$.
For each object, its representation $\textbf{e}_d$ is generated by the base model $\mathcal{B}$.
For each label, a label embedding $\textbf{l} \in \mathbb{R}^C$ is randomly initialized and updated during training.
The embeddings of the object $\textbf{e}_d$ and currently assigned label $\textbf{l}_t$ are concatenated and projected to a vector $\textbf{s}_t \in \mathbb{R}^{C}$ via a two-layer feed-forward network.
$\textbf{s}_t$ has the same size as the label embedding $\textbf{l}$ and is used as the state representation at time step $t$.
By stacking the action embeddings (\ie, the embeddings of candidate labels and \textit{stop} action), we obtain an action matrix $\textbf{A}_t$ with size $|\mathcal{A}| \times C$.
$\textbf{A}_t$ is multiplied with the state embedding $\textbf{s}_t$, which outputs the probability distribution of actions.
Finally, an action $a_{t}$ is sampled based on the probability distribution of the action space.
\vspace*{-.25cm}
\begin{equation*}
    \begin{split}
    \textbf{s}_t &= \text{ReLU}(\textbf{W}^1_l \text{ReLU}(\textbf{W}^2_l [\textbf{e}_d; \textbf{l}_t])),\\
    \pi(\textbf{a}_t\ |\ \textbf{s}; \textbf{W}) &= \text{softmax}(\textbf{A}_t \textbf{s}_t),\\
    a_t &\sim \pi(\textbf{a}_t\ |\ \textbf{s}; \textbf{W}).
\end{split}
\vspace*{-.25cm}
\end{equation*}
We use policy gradient~\citep{williams1992simple} as the optimization algorithm.
In addition, we adopt a \textit{self-critical} training approach~\citep{rennie2017self}. For each object $x_i$, two label assignments are generated: $\tilde{L}_{x_i}$ is sampled from the probability distribution, and $\hat{L}_{x_i}$, the baseline label assignment, is greedily obtained by choosing the action with the highest probability at each time step.
We use $\tilde{r}_t^{x_i} = r_t^{\tilde{L}_{x_i}} - r_t^{\hat{L}_{x_i}}$ as the actual reward, which ensures that the policy network learns to place the object to positions with higher F1 score than the greedy baseline.
Formally, we measure the global loss $\mathcal{O}_{g}$ as follows.
\vspace*{-.25cm}
\begin{equation*}
    \mathcal{O}_{g} =  -\sum_{i=1}^N \sum_{t=1}^T log \pi^{x_i}(a_t\ |\ \textbf{s}; \textbf{W}) \times v^{x_i}_t,
    \vspace*{-.25cm}
\end{equation*}
where $v^{x_i}_j = \sum_{t=j}^T \gamma^{t-j} \tilde{r}_t^{x_i}$ is the cumulative future reward at time $j$ and $\gamma \in [0, 1]$ is the discount factor.
At the time of inference, we greedily select labels with the highest probability as  $\hat{L}_{x_i}$.

\section{End-to-End Model Learning}
\subsection{Top-Down Supervised Pre-Training}
\label{subsec_tdsp}

Instead of learning from scratch, we use supervised learning to pre-train HiLAP.
We denote the supervised variant as HiLAP-SL.
While most parameters of HiLAP-SL are shared and used to initialize HiLAP (except that $\textbf{e}_{\text{stop}}$ is randomly initialized), its way of exploring the label hierarchy $H$ is dissimilar.

The major difference is that HiLAP-SL explores the label hierarchy $H$ in a top-down manner independently.
At each time step $t$, the object goes down one level on the hierarchy and the labels under the same parent are discriminated locally.
Specifically, the local per-parent label probability distribution $\textbf{p}_t^{\text{Local}}$ is estimated as $\textbf{p}_t^{\text{Local}} = \sigma (\textbf{C}_t \textbf{s}_t)$,
where $\sigma$ denotes the sigmoid function, and $\textbf{C}_t \in \mathbb{R}^{|\mathcal{C}(l_t)| \times C}$ denotes the candidate embeddings of HiLAP-SL, \ie, an embedding matrix consisting of the children of \textit{current} label $l_t$, rather than \textit{all} the labels where $x_i$ has been placed. 

Another difference is that in HiLAP the actions are sampled and thus might place the objects to incorrect labels, while in HiLAP-SL only the ground-truth positions are traversed during training.
Specifically, if there are $K (\geq 1)$ \textit{ground-truth} labels at the same level, the object embedding $\textbf{e}_d$ would be copied $K$ times and losses on the $K$ different paths would be measured independently (see Fig.~\ref{fig:example-sl} in Appendix for illustration).
The local loss of HiLAP-SL is defined as $\mathcal{O}_{l} =  \sum_{t=0}^T \mathcal{O}_t$,
where $T$ is the lowest label's level of one example and $\mathcal{O}_t$ estimates the binary cross entropy over the candidate labels $\mathcal{C}(l_t)$:
$\mathcal{O}_{t} =  -\sum_{i=1}^N \sum_{l \in \mathcal{C}(l_{t, i})}  L_i(l) \times  log \textbf{p}_{t, i}^{\text{Local}}(l) + (1 - L_i(l))  \times log(1 - \textbf{p}_{t, i}^{\text{Local}}(l))$,
where $L_i(l)$ and $\textbf{p}_{t, i}^{\text{Local}}(l)$ evaluate label $l$ of $x_i$.
Intuitively, HiLAP-SL works as if there were a set of local classifiers, although most of its parameters (except for the label embedding $\textbf{l}$) are shared by all the labels so that there is no need to train multiple classifiers.

\subsection{Combining Flat, Local, and Global Information for Policy Learning}
\label{subsec_allThreeLoss}
We further add a \textit{flat component} to HiLAP as a regularization of the base model.
Specifically, the flat component is a feed-forward network that projects the object embedding $\textbf{e}_d$ to a label probability distribution $\textbf{p}^{\text{Flat}}$ of all the labels on the hierarchy: $\textbf{p}^{\text{Flat}} = \sigma (\textbf{W}^{\text{Flat}} \textbf{e}_d)$.
The combination of the base model and the flat component functions the same as a flat model and ensures that the object representation $\textbf{e}_d$ has the capability of flat classification.
We denote the flat loss that measures the binary cross entropy over all the labels by
$\mathcal{O}_{f} =  -\sum_{i=1}^N \sum_{l \in L}  L_i(l) \times  log \textbf{p}_i^{\text{Flat}}(l) + (1 - L_i(l))  \times log(1 - \textbf{p}_i^{\text{Flat}}(l))$.
Combining the flat and local losses, the supervised loss in HiLAP-SL is defined as $\mathcal{O}_{\text{SL}} = \lambda \mathcal{O}_f + (1 - \lambda) \mathcal{O}_l$, where $\lambda \in [0, 1]$ is the mixing ratio.
Similar to \citet{celikyilmaz2018deep}, we also found that mixing a proportion of the supervised loss is beneficial to the learning process of HiLAP. Further combining the global information $\mathcal{O}_{g}$ (\ie, $\mathcal{O}_{\text{RL}}$), the total loss of HiLAP is defined as $\mathcal{O} =  \mathcal{O}_{\text{RL}} + \alpha \mathcal{O}_{\text{SL}}$, where $\alpha$ is a scaling factor accounting for the difference in magnitude between $\mathcal{O}_{\text{RL}}$ and $\mathcal{O}_{\text{SL}}$.
While we do not use the flat component during inference, it helps the representation learning of the base model and improves the performance of both HiLAP-SL and HiLAP (see Sec.~\ref{sec_ablation}).

\section{Experiments}

\subsection{Experiment Setup}
\start{Datasets.}
We conduct extensive experiments on five public datasets from various domains (summarized in Table~\ref{table_data} and detailed in Appendix~\ref{app_data_stat}).
The first two datasets are related to news categorization, including RCV1~\citep{lewis2004rcv1} and the NYT annotated corpus~\citep{sandhaus2008new}.
The third dataset is the Yelp Dataset Challenge 2018\footnote{\url{https://www.yelp.com/dataset/challenge}}.
We hypothesize that one business can be represented by its reviews and use the reviews to predict business categories.
The last two datasets are related to protein functional catalogue (FunCat) and gene ontology (GO) prediction~\cite{vens2008decision}, which are used to test the generalization ability of HiLAP to non-textual data.
For all the datasets, the lowest labels of one example may not be at the leaf nodes and there could be multiple labels at each level, making them harder and more realistic than \textit{mandatory-leaf} or \textit{single-path} datasets such as IPC~\cite{wipo2014international} and LSHTC~\cite{PartalasKBAPGAA15}.

\start{Evaluation Metrics.}
\label{subsec:metric}
We use standard metrics~\citep{johnson2014effective,Meng2018WeaklySupervisedNT,peng2018large} for HTC, including \textbf{Micro-F1}, \textbf{Macro-F1}, and \textbf{Example-based F1 (EBF)}~\cite{PartalasKBAPGAA15,peng2016deepmesh}.
Let $TP_i$, $FP_i$, $FN_i$ denote the true positive, false positive, and false negative for the i-th \textit{example} $x_i$ in object set $X$, respectively.
EBF calculates the F1 scores of all the \textit{examples} independently and averages them.
$P^{x_i} = \frac{TP_i}{TP_i + FP_i}$, $R^{x_i} = \frac{TP_i}{TP_i + FN_i}$, $\text{F1}^{x_i}=\frac{2P^{x_i} \times R^{x_i}}{P^{x_i} + R^{x_i}}$, and $\text{EBF} = \frac{1}{N} \sum_{i=1}^N \text{F1}^{x_i}$. 
Recall that $\text{F1}^{x_i}$ is used as the reward in HiLAP.

\begin{table}[t]
    \caption{\textbf{Statistics of the datasets.} $|L|$ denotes the number of labels in the label hierarchy. Avg($|L_i|$)  and Max($|L_i|$) denote the average and maximum number of labels of one object, respectively.} 
    \label{table_data}
    \vspace*{-.2cm}
    \centering
    \scalebox{.56}{
    \begin{tabular}{cccccccc}
        \toprule
      Dataset   & Hierarchy & $|L|$ & Avg($|L_i|$) & Max($|L_i|$) &Training & Validation &Test \\
        \midrule
         RCV1 & Tree & 103 & 3.24 & 17 & 23,149 & 2,315 &781,265 \\
         NYT & Tree & 115 & 2.52 & 14 & 25,279 & 2,528 & 10,828 \\
         Yelp & DAG & 539 & 3.77 & 32 & 87,375 & 8,737 & 37,265 \\
         FunCat & Tree & 499 & 8.76 & 45 & 1,628 & 848 & 1,281 \\
         GO & DAG & 4,125 & 34.9 & 141 & 1,625 & 848 & 1,278 \\
        \bottomrule
        \end{tabular}
    }
\vspace*{-.1cm}
\end{table}

\start{Base Models for Feature Encoding.}
Different from most of existing \textit{global} HTC methods that rely on pre-specified features~\citep{gopal2013recursive} as input or build on specific models~\citep{cai2004hierarchical,vens2008decision,silla2009global}, our framework is trained in an end-to-end manner by leveraging a differentiable feature representation learning model as the \textit{base model}.
Specifically, we use \textbf{TextCNN}~\citep{kim2014convolutional}, \textbf{HAN}~\citep{yang2016hierarchical}, \textbf{bow-CNN}~\citep{johnson2014effective} on the three textual datasets, and a \textbf{feed-forward network} on the two non-textual datasets.
The details of the base models are provided in Appendix~\ref{app_base} due to limited space.

To incorporate one base model into our framework, we remove its final feed-forward layer that projects the object representation $\textbf{e}_d$ to a flat probability distribution of all labels ($\textbf{p}^{\text{Flat}}$), and use $\textbf{e}_d$ directly as the input of HiLAP. As one will see in the later experiments, HiLAP consistently improves the base model by modeling the label hierarchy in an effective manner.

\begin{table}[t]
    \caption{\textbf{Performance comparison on RCV1.} * denotes the results reported in~\citet{peng2018large} on the same dataset split. Note that the results of HR-SVM reported in~\citet{gopal2013recursive} are not comparable as they use a different hierarchy with 137 labels. } 
    \vspace{-.1cm}
    \label{table_rcv1}
    \centering
    \scalebox{.83}{
    \begin{tabular}{clcccc}
        \cmidrule[0.06em]{2-5}
      & \textbf{Method}  & Micro-F1 & Macro-F1 & EBF \\
       \cmidrule{2-5}
        \multirow{5}{*}{\rotatebox[origin=c]{90}{Flat}} &$\text{Leaf-SVM}^*$ & 69.1 & 33.0 & - \\
        &SVM & 80.4 & 46.2 & 80.5 \\
        &TextCNN & 76.6 & 43.0 & 75.8 \\
        &HAN & 75.3 & 40.6 & 76.1 \\
        &bow-CNN & 82.7 & 44.7 & 83.3\\
        \cmidrule{2-5}
        \multirow{8}{*}{\rotatebox[origin=c]{90}{Local \& Global}}&$\text{TD-SVM}$& 80.1 & 50.7 & 80.5 \\
        &$\text{HSVM}^*$ & 69.3 & 33.3 & - \\
        &$\text{HR-SVM}^*$ & 72.8 & 38.6 & - \\
        &$\text{HR-DGCNN}^*$ & 76.1 & 43.2 & - \\
        &HMCN & 80.8 & 54.6 & 82.2 \\
        
        &\textbf{HiLAP} (TextCNN) & 78.6 & 50.5 & 80.1 \\
        &\textbf{HiLAP} (HAN)  & 75.4 & 45.5 & 77.4 \\
        &\textbf{HiLAP} (bow-CNN)  & \textbf{83.3} & \textbf{60.1} & \textbf{85.0} \\
        \cmidrule[0.06em]{2-5}
        \end{tabular}
    }
\vspace*{-.1cm}
\end{table}

\subsection{Compared Methods}
\start{1. Traditional HTC Methods.}
A major line of work for HTC is Support Vector Machines (SVM) and its hierarchical variants.
Specifically, \textbf{SVM} performs standard multi-label classification using one-vs-the-rest (OvR) strategy.
\textbf{Leaf-SVM} treats each leaf node as a label and adds the ancestors of predicted leaf nodes. 
Variants such as \textbf{HSVM}~\citep{tsochantaridis2005large}, Top-Down SVM (\textbf{TD-SVM})~\citep{liu2005support}, and Hierarchically Regularized SVM (\textbf{HR-SVM})~\citep{gopal2013recursive} are also tested.
Other state-of-the-art HTC methods that we compare with include \textbf{Clus-HMC}~\cite{vens2008decision} and \textbf{CSSA}~\cite{bi2011multi}.

\start{2. Neural HTC Methods.}
There are not many neural methods that specifically target HTC. 
We mainly compare with two \textit{latest} neural models: \textbf{HR-DGCNN}~\citep{peng2018large}, which extends hierarchical regularization~\citep{gopal2013recursive} to Graph-CNN and compares favorably to flat models like RCNN~\cite{lai2015recurrent} and XML-CNN~\cite{liu2017deep}, and \textbf{HMCN}~\citep{wehrmann2018hierarchical}, which outperforms state-of-the-art HTC methods such as HMC-LMLP~\cite{cerri2016reduction}.
We also compare with the base models that we use for feature encoding.
The main aim is to see how much gain they could obtain by combining each one of them with HiLAP.

\subsection{Implementation Details}
For datasets without held-out set, we randomly sample 10\% from the training set as the validation set following~\citet{johnson2014effective,peng2018large}.
We only use the first 256 tokens of each document for representation learning.
All the models are trained using an Adam optimizer with initial learning rate 1e-3 and weight decay 1e-6.
We use GloVe~\citep{pennington2014glove} with size 50 as word embeddings for TextCNN~\cite{kim2014convolutional} and HAN~\cite{yang2016hierarchical}.
We create a vocabulary of the most frequent 30,000 words in the training data and generate multi-hot vectors as the input of bow-CNN~\cite{johnson2014effective}.
For our framework, since the parameter updates are performed after $T$ steps, we cache the object representation $\textbf{e}_d$ and reuse it at each step for better efficiency.
More details are provided in Appendix~\ref{app_implementation} for reproducibility.

\begin{table}[t]
    \caption{\textbf{Performance comparison on the NYT and Yelp datasets.} We mainly compare with competitive baselines that perform well on RCV1.}
    \label{table_yelp}
    \vspace{-.1cm}
    \centering
    \scalebox{.58}{
    \begin{tabular}{lcccccc}
        \cmidrule[0.08em]{1-7}
       \multirow{2}{*}{\textbf{Method}}   &  & \textbf{NYT} &  &  & \textbf{Yelp} & \\
      \cmidrule(r){2-4} \cmidrule(r){5-7}
       & Micro-F1 & Macro-F1 & EBF & Micro-F1 & Macro-F1 & EBF\\
        \cmidrule{1-7}
        SVM  & 72.4 & 37.1 & 74.0 & 66.9 & 36.3 & 68.0\\
        TextCNN  & 69.5 & 39.5 & 71.6 & 62.8 & 27.3 & 63.1\\
        HAN  & 62.8 & 22.8 & 65.5 & 66.7 & 29.0 & 67.9\\
        bow-CNN& 72.9 & 33.4 & 74.1 & 63.6 & 23.9 & 63.9\\
        TD-SVM & 73.7 & 43.7 & 75.0 & 67.2 & 40.5 & 67.8  \\
        HMCN& 72.2 & 47.4 & 74.2 & 66.4 & 42.7 & 67.6 \\
        \cmidrule{1-7}
        \textbf{HiLAP} (TextCNN) & 69.9 & 43.2 & 72.8  & 65.5 & 37.3 & 68.4 \\
        \textbf{HiLAP}  (HAN)  & 65.2 & 28.7 & 68.0 & \textbf{69.7} & 38.1 & \textbf{72.4} \\
        \textbf{HiLAP}  (bow-CNN) & \textbf{74.6} & \textbf{51.6} & \textbf{76.6} & 68.9 & \textbf{42.8} & 71.5 \\
        \cmidrule[0.08em]{1-7}
        
        \end{tabular}
    }
    \vspace*{-.1cm}
\end{table}

\begin{figure*}[t]
    \centering
    \includegraphics[width=0.97\linewidth]{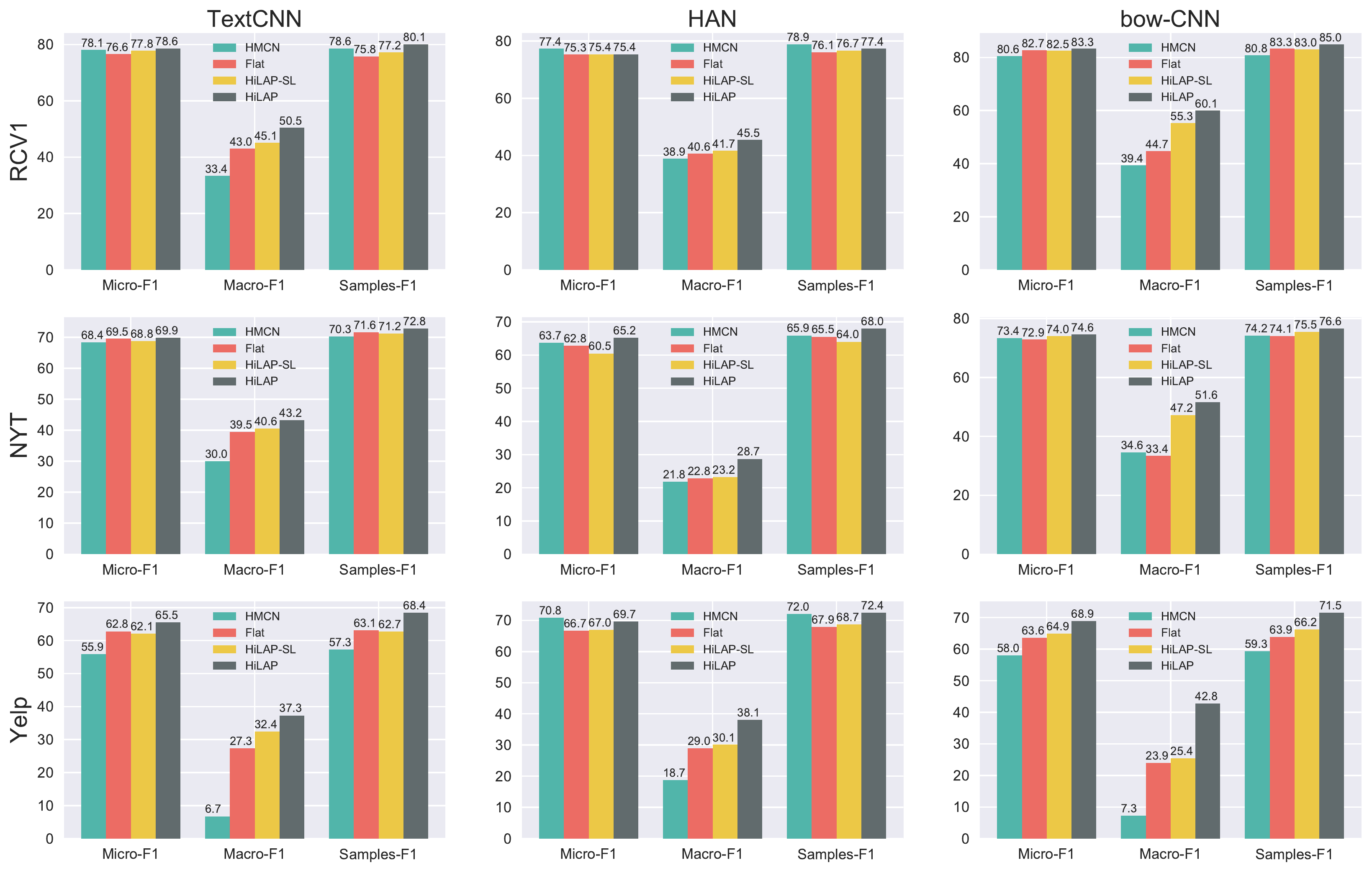}
    \vspace{-.2cm}
     \caption{\textbf{Performance comparison of different classification frameworks using the same base models.} We compare HiLAP with its flat, supervised variants, and HMCN. Results show that HiLAP exhibits consistent improvement over flat classifiers and larger gains than HMCN.}
    \label{fig:bar}
    \vspace*{-.2cm}
\end{figure*}

\subsection{Performance Comparison}
\noindent
\textbf{1. Comparison with State-of-the-art Methods.}
We compare the performance of HiLAP to state-of-the-art HTC methods and show the results in Tables~\ref{table_rcv1} and~\ref{table_yelp}.
On RCV1, HiLAP (HAN) achieves similar performance to HR-DGCNN even though the corresponding base model HAN is originally worse than HR-DGCNN.
HiLAP (TextCNN) outperforms most baselines in Macro-F1 and perform similarly to TD-SVM despite that it uses one global classifier while TD-SVM uses a set of classifiers.
Among all compared methods, HiLAP (bow-CNN) achieves the best performance on all the three metrics.\footnote{The results are not comparable with~\citet{johnson2014effective} due to implementation details and the fact that they tune the threshold for each label using k-fold cross-validation. See Appendix~\ref{app_baseline} for more discussions.}
On NYT, similar results are observed:
TextCNN and HAN are both improved when combining with HiLAP and
HiLAP (bow-CNN) again achieves the best performance.
On Yelp, HiLAP (HAN) achieves the best Micro-F1 and EBF, while HiLAP (bow-CNN) obtains the highest Macro-F1.

\smallskip
\noindent
\textbf{2. Comparison using Same Base Models.}
We compare the performance of different frameworks that support the use of \textit{exactly the same} base models and summarize the results in Fig.~\ref{fig:bar}.\footnote{For HMCN, we replace its static features with the same base model for fair comparison.}
Due to the extreme imbalance of the data, directly applying a flat model may suffer from low Macro-F1, \ie, the predictions of flat models are inevitably biased to the most popular labels.
HMCN also has the same issue, resulting in Macro-F1 lower than 10 when combining with some base models.
In contrast, HiLAP outperforms the baselines significantly in Macro-F1, which implies that our method is better at tackling labels with relatively few examples.
It is also observed that HiLAP-SL sometimes may have a negative effect in terms of Micro-F1, although it is usually marginal compared with the gains in Macro-F1.
However, such negative effects are eliminated by HiLAP through better exploration of the label hierarchy.
Overall, HiLAP achieves the highest performance on 24 of 27 results among the combinations of three datasets, three base models, and three evaluation metrics.
In particular, HiLAP yields an average improvement of 33.4\% in Macro-F1 compared to corresponding base models.

\begin{table}[t]
    \caption{\textbf{Performance comparison on Functional Catalogue and Gene Ontology.} We compare with state-of-the-art hierarchical classification methods that take exactly the same raw features as input (\ie, we exclude models designed specifically for text objects).} 
    \label{table_fungo}
    \centering
        \vspace{-.1cm}
    \scalebox{.63}{
        \begin{tabular}{lcccccc}
            \cmidrule[0.08em]{1-7}
          \multirow{2}{*}{\textbf{Method}}   & & \textbf{FunCat} &  &  & \textbf{GO} & \\
          \cmidrule(r){2-4} \cmidrule(r){5-7}
          & Micro-F1 & Macro-F1 & EBF & Micro-F1 & Macro-F1 & EBF\\
            \cmidrule{1-7}
            SVM  & 2.72 & 1.21 & 3.42 & 34.1 & 1.46 & 36.8\\
            CSSA  & 16.0 & 4.60 & 14.8 & 11.6 & 0.76 & 11.5\\
            CLUS-HMC & 25.2 & 4.14 & 24.1 & 41.4 & 3.01 & 40.3\\
            HMCN & 21.3 & 5.07 & 21.5 & 43.2 & 3.81 & 43.3\\
            \textbf{HiLAP} & \textbf{26.5} & \textbf{7.50} & \textbf{27.4} & \textbf{45.4} & \textbf{5.87} & \textbf{45.2}\\

            \cmidrule[0.08em]{1-7}
        
        \end{tabular}
    }
\vspace*{-.1cm}
\end{table}

\smallskip
\noindent
\textbf{3. Results on Functional Genomics Prediction.}
We compare HiLAP with CSSA~\cite{bi2011multi}, CLUS-HMC~\citep{vens2008decision}, and HMCN~\citep{wehrmann2018hierarchical} on the FunCat and GO datasets, as they represent the state-of-the-art on these datasets.
An SVM classifier is also evaluated to better understand the difficulties of the task.
We use the same raw features as the input of all the methods for apples-to-apples comparison and list the results in Table~\ref{table_fungo}.
Note that the metric \textit{area under the average precision-recall curve (AUPRC)}~\cite{wehrmann2018hierarchical} is not applicable because HiLAP does not use a flat probability distribution of all the labels.
As one can see, HiLAP outperforms all the baselines on both datasets by a large margin.
In particular, we observe significant improvement on Macro-F1 over the best baseline (47.9\% and 53.9\%, respectively), which shows that our method is especially better at classifying sparse labels than previous approaches.

\subsection{Performance Analysis}
\label{sec_ablation}
\begin{table}[t]
    \caption{\textbf{Ablation study of HiLAP.} We evaluate variants of HiLAP using bow-CNN~\citep{johnson2014effective} on RCV1~\citep{lewis2004rcv1}.}
    \label{table_ablation}
    \centering
        \vspace{-.1cm}
    \scalebox{.8}{
    \begin{tabular}{lccc}
        \toprule
      Method   & Micro-F1 & Macro-F1 & EBF \\
        \midrule
         Flat-Only & 82.7 & 44.7 & 83.3\\
         HiLAP-SL-NoFlat & 81.0 & 52.1 & 81.7\\
        HiLAP-SL & 82.5 & 55.3 & 83.0\\
         HiLAP-NoSL & 83.2 & 59.3 & 85.0\\
         HiLAP-NoFlat & 83.0 & 59.8 & 84.7\\
         HiLAP & \textbf{83.3} & \textbf{60.1} & \textbf{85.0}\\
        \bottomrule
        \end{tabular}
    }
\vspace*{-.1cm}
\end{table}
\start{1. Ablation Study on Different Framework Components.}
We show the ablation analysis of HiLAP in Table~\ref{table_ablation}.
Using \textit{Flat-Only} degenerates HiLAP to the flat baseline. By comparing the results of \textit{Flat-Only} and HiLAP-SL-NoFlat (a variant of HiLAP-SL without flat loss), we further confirm that flat approaches are likely to neglect sparse labels, which results in low Macro-F1. Local approaches (HiLAP-SL-NoFlat), on the other hand, are slightly worse in terms of Micro-F1 and EBF but significantly better on Macro-F1.
By combining flat and local information, HiLAP-SL achieves performance close to \textit{Flat-Only} on Micro-F1 and EBF, and even higher Macro-F1 than HiLAP-SL-NoFlat.
HiLAP-NoSL is initialized by the pre-trained HiLAP-SL model without mixing the supervised loss during its training.
We can see that using the reinforced loss alone still improves the performance on all the three metrics.
After removing the flat loss during the training of HiLAP, HiLAP-NoFlat shows slightly lower performance than the full HiLAP model, indicating that the flat component serves as a regularization of the base model and is beneficial to the overall performance.

\begin{figure}[ht]
    \centering
    \includegraphics[width=1.02\linewidth]{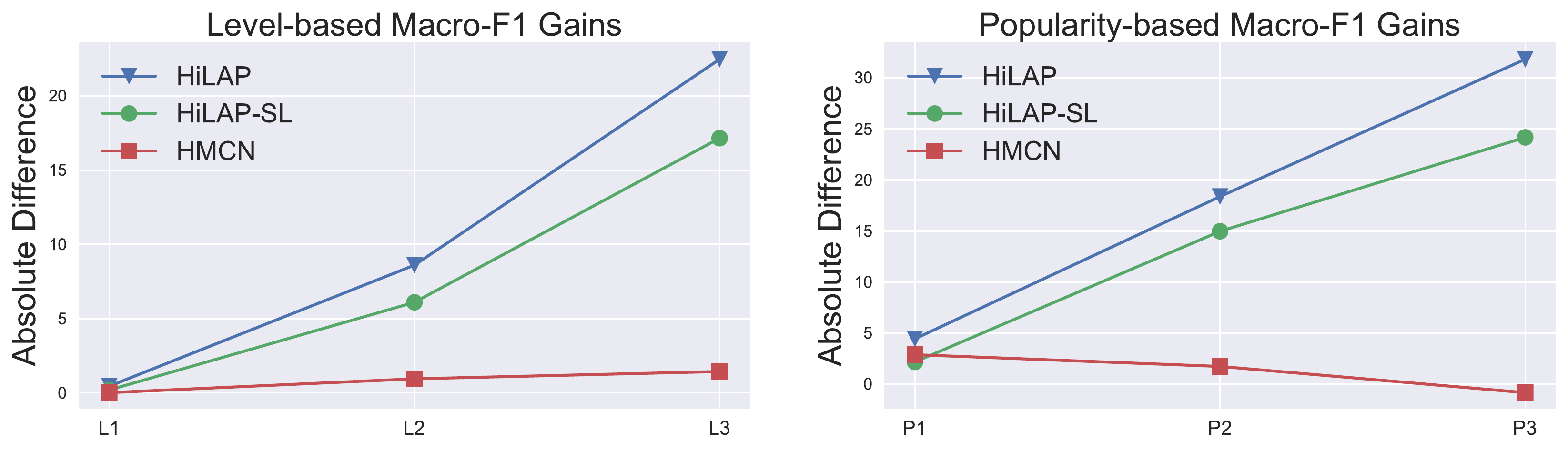}
    \vspace{-.1cm}
     \caption{\textbf{Performance Study on Label Granularity and Popularity.} We compute level-based and popularity-based Macro-F1 gains on NYT with bow-CNN as base model. We denote the levels of the hierarchy with L1, L2, and L3 (\textit{left}) and divide the labels into three equal sized categories (P1, P2, and P3) in a descending order by their number of examples (\textit{right}).}
    \label{fig:perlevel}
    \vspace{-.15cm}
\end{figure}

\start{2. Performance Study on Label Granularity and Popularity.}
We analyze the sources of performance gains by dividing the labels based on their levels and number of supporting examples.
Fig.~\ref{fig:perlevel} shows the absolute Macro-F1 differences between several methods and the base model.
We observe similar results for other setups and omit them for a clearer view. As depicted in Fig.~\ref{fig:perlevel}, HiLAP and HiLAP-SL are especially beneficial to unpopular labels (P3) at the bottom levels (L3).

\start{3. Analysis of Label Inconsistency.}
Label inconsistencies often happen in approaches that perform flat inference, but they are not measured by standard evaluation metrics like F1 scores.
To provide a picture of how severe the issue is, we further conduct experiments to check the percentage of objects that are predicted with inconsistent labels (Table~\ref{table_inconsistencies}).
We found, for example, 29,186/781,265 (3.74\%) predictions of TextCNN have inconsistent on RCV1. 
In contrast, HiLAP ensures \textbf{0\%} label inconsistency without the need of post-processing, because its predictions are always valid sub-trees of the label hierarchy (refer to Fig.~\ref{fig:example}). 

\begin{table}[th]
    \caption{\textbf{Analysis of Label Inconsistency.} We compare various methods by the percentage of predictions with inconsistent labels on RCV1~\citep{lewis2004rcv1}.} 
    \label{table_inconsistencies}
    \centering
        \vspace{-.1cm}
    \scalebox{.75}{
    \begin{tabular}{lccc}
        \toprule
       SVM  & TextCNN & HMCN & HiLAP \\
        \midrule
         4.83\% & 3.74\% & 3.84\% & \textbf{0\%} \\
        \bottomrule
        \end{tabular}
    }
\vspace*{-.1cm}
\end{table}

\section{Related Work}
Hierarchical classification approaches have been developed for many applications.
For text classification, both traditional methods~\citep{lewis2004rcv1,gopal2013recursive} and neural methods~\citep{johnson2014effective,peng2018large} have been proposed to classify, \eg, the topics of newswire and web content~\citep{Sun2001HierarchicalTC} or categories of laws and patents~\citep{bi2015bayes,cai2004hierarchical,rousu2005learning}.
Many previous studies~\citep{liu2005support,Sun2001HierarchicalTC} train a set of local classifiers and make predictions in a top-down manner. In particular, \citet{bi2015bayes} develop Bayes-optimal predictions that minimize the global risks but their model is still locally trained.
Such local approaches are not popularly used among recent neural-based HTC models~\citep{johnson2014effective,peng2018large} since it is usually infeasible to train many neural classifiers locally.

Global methods, on the other hand, train only one classifier.
Although global methods are desirable, they are relatively less studied due to the complexity of the problem.
Existing global models are generally modified based on specific flat models.
Hierarchical-SVM \citep{cai2004hierarchical,qiu2009hierarchical} generalizes Support Vector Machine (SVM) learning based on discriminant functions that are structured in a way that mirrors the label hierarchy. One limitation is that Hierarchical-SVM only supports balanced tree (all possible labels are presumed to be at the same level in their experiments).
Hierarchical naive Bayes~\citep{silla2009global} modifies naive Bayes by updating weights of one's ancestors as well whenever one label's weights are updated.
There are other global methods that are based on association rules~\citep{wang2001hierarchical}, C4.5~\citep{clare2003predicting}, kernel machines~\citep{rousu2005learning}, and decision tree~\citep{vens2008decision}.
Constraints such as the regularization that enforces the parameters of one node and its parent to be similar~\citep{gopal2013recursive} are also proposed to leverage the label hierarchy while maintaining scalability. However, their use of the label hierarchies is somewhat limited compared with HiLAP.

\section{Conclusions}
We proposed an end-to-end reinforcement learning approach to hierarchical text classification (HTC) where objects are labeled by placing them at the proper positions in the label hierarchy.
The proposed framework makes \textit{consistent} and \textit{inter-dependent} predictions, in which any neural-based representation learning model can be used as a base model and a label assignment policy is learned to determine where to place the objects and when to stop the assignment process.
Experiments on five public datasets and four base models showed that our approach outperforms state-of-the-art HTC methods significantly.
For future work, we will explore the effectiveness of the proposed framework on other base models and forms of data (\eg, images).
We will introduce more losses covering other aspects in the objective function to further improve the performance of our framework.

\section*{Acknowledgments}
 Research was sponsored in part by U.S. Army Research Lab under Cooperative Agreement No. W911NF-09-2-0053 (NSCTA), DARPA under Agreement No. W911NF-17-C-0099, National Science Foundation IIS 16-18481, IIS 17-04532, and IIS-17-41317, grant 1U54GM-114838 awarded by NIGMS, National Science Foundation SMA 18-29268, DARPA MCS and GAILA, IARPA BETTER, Schmidt Family Foundation, Amazon Faculty Award, Google Research Award, Snapchat Gift, and JP Morgan AI Research Award.
 We thank Chao Zhang, Xiao-Yang Liu, Qingrong Chen, Jun Yan, collaborators in the INK research lab, and anonymous reviewers for their help and valuable feedback.

\bibliography{emnlp-ijcnlp-2019}
\bibliographystyle{acl_natbib}

\appendix

\clearpage
\section{Reproducibility Details of Datasets}
\label{app_data_stat}
In this section, we describe the details of the datasets used in our experiments.

The RCV1 dataset~\citep{lewis2004rcv1} is a manually labeled newswire collection of Reuters News from 1996 to 1997. 
Its news documents are categorized with three aspects: industries, topics, and regions.
We follow the original training/test split for RCV1 and use its topic-based label hierarchy for classification as it has been well used in prior work~\citep{gopal2013recursive,johnson2014effective,peng2018large,wehrmann2018hierarchical}.
There are 103 categories and four levels in total including all labels except for the root label in the hierarchy.

The NYT annotated corpus~\citep{sandhaus2008new} is a collection of New York Times news from 1987 to 2007.
Due to its large size, we randomly sampled 36,107 documents from all the news documents, and further split them into training and test set of 25,279 and 10,828 examples, respectively. 
We use the first three levels in the hierarchy and keep the labels with at least 40 supporting examples.

For the Yelp dataset, the label hierarchy is taken from the Yelp Business Categories\footnote{\url{https://www.yelp.com/developers/documentation/v3/all\_category\_list}}, which Fig.~\ref{fig:example} is a subset of.
For preprocessing, we first removed categories that have fewer than 100 businesses and then businesses that have fewer than 5 reviews.
We concatenated (at most) the first 10 reviews of each business as its representation.
We set the training/test ratio to 70\%/30\%, which results in a training set of 87,375 examples and a test set of 37,517 examples.
This is an even more challenging task because the reviews are usually written in an informal way and it is more imbalanced than the RCV1 or NYT datasets.
For example, label \textit{Restaurants} has 32,357 businesses in the training set while \textit{Retirement Homes} has 23.

For the FunCat and GO datasets, we take the \textit{cellcycle} data from~\cite{vens2008decision}\footnote{\url{https://dtai.cs.kuleuven.be/clus/hmcdatasets/}}.
Compared with the text datasets above, raw features are provided as input for all compared methods.
Furthermore, their training data is rather limited while the label space is much larger (4,125 vs. 539).
Since there are many labels that do not have any example in either training set or test set, we exclude such labels when calculating Macro-F1. Note that it does not have any effect on the ratio of results from two different methods as the F1 scores of those labels without supporting examples are always zero.
The features provided by the datasets are taken as input as they are except that the missing values are replaced with the mean value of corresponding features.
All the compared methods take the same raw features for fair comparison.

\section{Performance Analysis of Baselines}
\label{app_baseline}
There are several things to note in terms of the performance of the baselines.
First, our results are not comparable to~\citet{lewis2004rcv1,johnson2014effective} due to implementation details (\eg, we only take the first 256 tokens) and the fact that they tune the threshold for each label using \textit{scutfbr}~\cite{lewis2004rcv1}.
According to the implementation in LibSVM\footnote{\url{https://www.csie.ntu.edu.tw/~cjlin/libsvmtools/multilabel/}}, the \textit{scutfbr} threshold tuning algorithm uses two nested 3-fold cross validation for each of the 103 labels and the classifier is trained $3 \times 3 \times 103 = 927$ times, which is infeasible in our case.

Secondly, we found that the original performance of HMCN~\citep{wehrmann2018hierarchical} is sometimes much lower than expected.
After tuning their model, we observed that if we first conduct a weighted sum of the local and global outputs and then apply the sigmoid function, the performance of HMCN becomes much better (see Table~\ref{table_hmcn}) than doing them in the opposite order as in~\citet{wehrmann2018hierarchical}.
In addition, we found that HMCN + HAN~\citep{yang2016hierarchical} would result in extremely low performance. We had to remove HMCN's batch normalization to make it compatible with HAN.
Combining HMCN with other base models did not encounter similar issues.

\begin{table*}[t]
    \caption{Comparison of different implementations of HMCN.} 
    \label{table_hmcn}
    \centering
    \scalebox{.8}{
    \begin{tabular}{lccccccccc}
        \toprule
      \multirow{2}{*}{\textbf{Model}}   &  & \textbf{RCV1} &  &  & \textbf{Yelp} &  &  & \textbf{NYT} & \\
      & Micro-F1 & Macro-F1 & EBF & Micro-F1 & Macro-F1 & EBF & Micro-F1 & Macro-F1 & EBF\\
        \midrule
        HMCN (original) & 78.2 & 33.2 & 78.9 & 56.3 & 8.5 & 57.3 & 62.1 & 32.4 & 62.7 \\
        HMCN (ours)& 80.8 & 54.6 & 82.2 & 66.4 & 42.7 & 67.6 & 72.2 & 47.4 & 74.2\\
        \bottomrule
        
        \end{tabular}
    }
\end{table*}

Thirdly, our implementation of TextCNN~\citep{kim2014convolutional} and HAN~\citep{yang2016hierarchical} shows better performance than those reported in~\citet{peng2018large} due to implementation details. A comparison can be found in Table~\ref{table_www}.

\begin{table}[t]
    \caption{Comparison of different implementations of HAN and TextCNN on the RCV1 dataset.} 
    \label{table_www}
    \centering
    \scalebox{.8}{
    \begin{tabular}{lcc}
        \toprule
      Model  & Micro-F1 & Macro-F1 \\
        \midrule
        TextCNN (in~\citet{peng2018large}) & 73.2 & 39.9 \\
        TextCNN (ours)& 76.6 & 43.0\\
        HAN (in~\citet{peng2018large}) & 69.6 & 32.7\\
        HAN (ours)& 75.3 & 40.6\\
        \bottomrule
        
        \end{tabular}
    }
\end{table}

\section{Details of Base Models}
\label{app_base}
\smallskip
\noindent
\textbf{1. Base Models for Encoding Text Objects.}
For the text classification datasets, three representative text encoding models with different characteristics are selected as the base models to prove the robustness and versatility of HiLAP.
We briefly describe the base models and the reasons we choose them as follows.

\textbf{TextCNN}~\citep{kim2014convolutional} is a classic convolutional neural network for text classification.
    In our implementation, TextCNN is composed of one convolutional layer with three kernels of different sizes (3, 4, 5), followed by max pooling, dropout, and fully-connected layers.
    We choose TextCNN because it is one of the first successful and well used neural-based models for text classification.

\textbf{HAN}~\citep{yang2016hierarchical}
    first learns the representation of sentences by feeding words in each sentence to a GRU-based sequence encoder~\citep{bahdanau2014neural} and then feeds the representation of the encoded sentences into another GRU-based sequence encoder, which generates the representation of the whole document.
    Attention mechanism such as word attention and sentence attention is also used.
    We choose HAN because it uses RNNs instead of CNNs and is shown to be effective on the \textit{flat} Yelp Review datasets.

\textbf{bow-CNN}~\citep{johnson2014effective} employs bag of words (multi-hot zero-one vectors) as input to represent text objects and directly applies CNNs to the high-dimensional multi-hot vectors encoding. It learns the representation of small text regions (rather than single words) for use in classification. We choose bow-CNN since it does not use any word embeddings as in TextCNN and HAN. In addition, bow-CNN achieved state-of-the-art performance RCV1~\citep{lewis2004rcv1}.

\smallskip
\noindent
\textbf{2. Base Model for Encoding Raw Features.}
For functional genomics prediction, one feed-forward neural network is used for simplicity as raw features are already provided in the datasets. 

\section{Reproducibility Details of Implementation}
\label{app_implementation}
We implement the base models and HMCN~\cite{wehrmann2018hierarchical} according to the original papers and existing implementations.
We use the official implementation of Clus-HMC~\cite{vens2008decision}\footnote{\url{https://dtai.cs.kuleuven.be/clus/}} and one open-source implementation of CSSA~\cite{bi2011multi}\footnote{\url{https://github.com/sushobhannayak/cssag}}.
We use \textit{scikit-learn} for SVM-based methods.
TF-IDF features are used for text classification when raw features are needed as input.

For our framework, we specify the number of steps in HiLAP-SL to be the number of levels in the label hierarchy.
We set the maximum number of steps in HiLAP to be reasonably large (depending on the average number of labels of one object) so that it could explore the hierarchy and learn when to stop by itself.
For the purpose of batch training, we convert the original indefinite-horizon MDPs to finite-horizon by adding an absorbing state, \ie, after visiting the most fine-grained label in HiLAP-SL or entering the \textit{stop} state in HiLAP, it would loop in the current state until the maximum number of steps, waiting for other objects in the same batch to finish.

We set the size of $\textbf{W}^2_l$ to 500 and the sizes of $\textbf{W}^1_l$ and label embedding $\textbf{l}_t$ to 50 in all the text classification datasets and set them to 1,000 in the other datasets.
We did not observe clear performance changes when varying the probability of dropout in base models like TextCNN.
We set batch size to 32 as it performs well on the validation set and a batch size as large as 128 may cause performance losses.

\section{Additional Figure Illustration}

\begin{figure*}[th]
    \centering
    \includegraphics[width=1.0\linewidth]{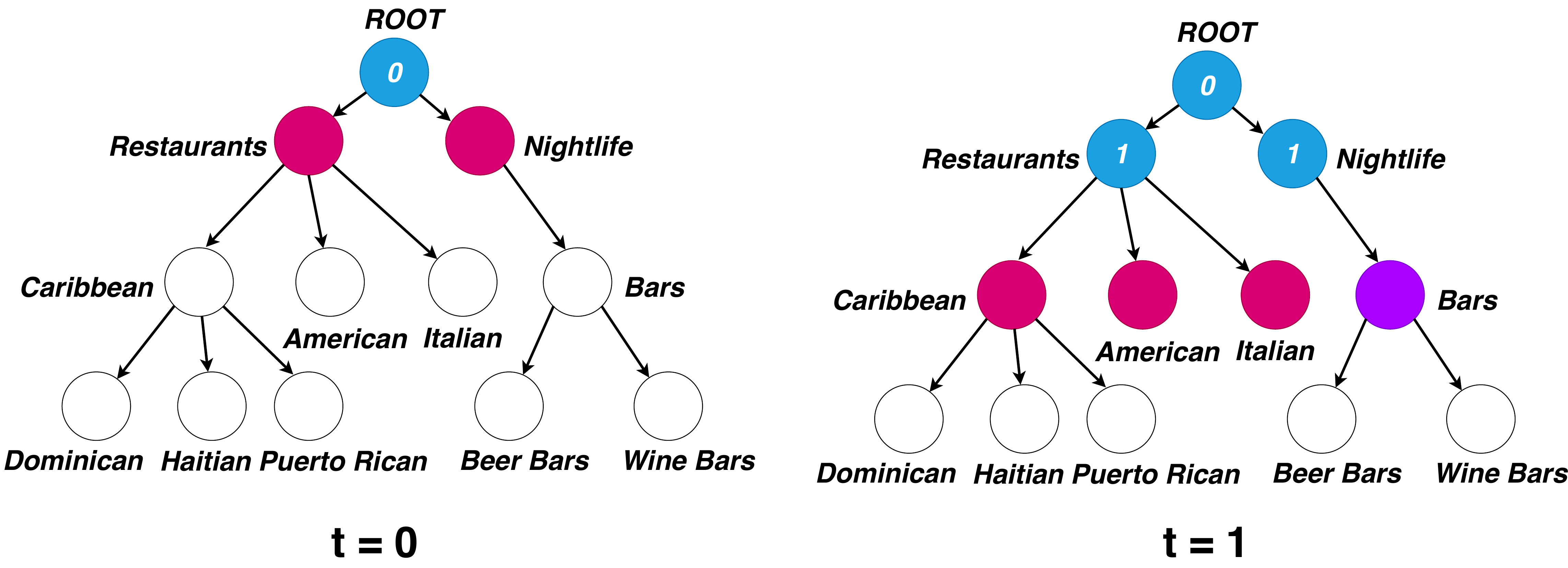}
    \vspace{-.3cm}
     \caption{\textbf{One time step in HiLAP-SL.} At $t=1$, two ($K=2$) local per-parent probabilities $\textbf{p}_1^{\text{Local}}$ are measured independently and aggregated in the loss function $\mathcal{O}_{1}$.}
    \label{fig:example-sl}
    \vspace{-.3cm}
\end{figure*}

\end{document}